\documentclass[preprintnumbers,amsmath,amssymb,floatfix,superscriptaddress,nofootinbib,12pt]{revtex4}
\usepackage{graphicx}
\usepackage{epsfig}
\usepackage{bm}
\usepackage{amsfonts}
\def\be {\begin{equation}}
\def\ee {\end{equation}}
\def\ba {\begin{eqnarray}}
\def\ea {\end{eqnarray}}
\begin{document}

\title{Black Holes in Bulk Viscous Cosmology}

\author{Francesco De Paolis}
\email{francesco.depaolis@le.infn.it} \affiliation{Department of Physics, University of Lecce, via
Arnesano, I-73100, Lecce, Italy}
\author{Mubasher Jamil}
\email{mjamil@camp.nust.edu.pk} \affiliation{Center for Advanced
Mathematics and Physics,\\ National University of Sciences and
Technology, Rawalpindi, 46000, Pakistan}
\author{Asghar Qadir}
\email{aqadirmath@yahoo.com} \affiliation{Center for Advanced
Mathematics and Physics,\\ National University of Sciences and
Technology, Rawalpindi, 46000, Pakistan}
\begin{abstract}
\textbf{Abstract:} We investigate the effects of the accretion of phantom energy with
non-zero bulk viscosity onto a Schwarzschild black hole and show
that black holes accreting viscous phantom energy will lose mass
rapidly compared to the non-viscous case. When matter is
incorporated along with the phantom energy, the black holes meet
with the same fate as bulk viscous forces dominate matter accretion.
If the phantom energy has large bulk viscosity, then the mass of the
black hole will reduce faster than in the small viscosity case.
\end{abstract}
\maketitle
\textbf{Keywords}: Accretion; black hole; bulk
viscosity; phantom energy.
\newpage
\large
\section{Introduction}

Observations of WMAP \cite{Benn,wang,sper} and supernova of type Ia
data \cite{perl} have revealed that our Universe is filled with an
exotic dark energy apart from dark matter. The nature and
composition of this energy is still an open problem but its dynamics
is well understood i.e it causes an approximately exponential
expansion of the Universe (see \cite{pad} for recent reviews on
dark energy). Astrophysical data suggest that about two thirds of
the critical energy density is stored in the dark energy component.
For the equation of state (EoS) parameter $\omega<-1$, the fluid is called phantom
energy (PE). Observations show that  $\omega$ is
constrained in the range $-1.38<\omega<-0.82$ \cite{Melch}, thus providing evidence of phantom energy in the Universe.
The PE violates all the energy conditions in all forms
(weak, null, strong or dominant). The phantom energy
can cause some peculiar phenomena e.g. the existence of wormholes
\cite{kuhf,rahm}, infinite expansion of the Universe in a finite
time causing a Big Rip (BR) and the destruction of all
gravitationally bound structures including black holes
\cite{cald,babi,ness,mota}. In particular, black holes will
continuously lose mass and disappear near the BR
(see \cite{rahm1,svet} for the opposite viewpoint).

Dark energy with bulk viscosity has a peculiar
property to cause accelerated expansion of phantom type in the late
evolution of the universe \cite{brevik}. It can also alleviate
several cosmological puzzles like cosmic age problem \cite{ricci},
coincidence problem \cite{problem} and phantom crossing \cite{ivor}. We will consider phantom energy as an imperfect fluid,
implying that the PE could contain non-zero bulk and
shear viscosities \cite{coles}. The bulk viscosities are negligible
for non-relativistic and ultra-relativistic fluids but are important
for the intermediate cases. In viscous cosmology, shear viscosities
arise in relation to space anisotropy while the bulk viscosity
accounts for the space isotropy \cite{brevik,hu}. Generally, shear
viscosities are ignored (as the CMB does not indicate significant
anisotropies) and only bulk viscosities are taken into account for
the fluids in the cosmological context. Moreover, bulk viscosity
related to a grand unified theory phase transition may lead to an
explanation of the accelerated cosmic expansion \cite{lang}.

Babichev et al \cite{babi} studied the effects of the accretion
of phantom energy onto a Schwarzschild black hole taking PE to be a
perfect fluid. As a first approximation, the bulk viscosity can be
ignored, but to get a better picture we need to incorporate it into
the phantom fluid. We have adopted the procedure of \cite{babi,Mich}
for our calculations.

The plan of the paper is as follows: in the next section we
review viscous cosmology; in section three we
discuss the relativistic model of accretion onto a black hole; in
the subsequent section we use results from viscous cosmology for the
accretion model; next we give two examples to illustrate the
accretion process with a constant and power law viscosity. In
section six we study black hole evolution in the presence of matter
and viscous phantom energy. Finally we conclude the paper with a brief
discussion of our results.

\section{Bulk-viscous cosmology}

We assume the
background spacetime to be homogeneous, isotropic and spatially flat
($k=0$) and described by the Friedmann-Robertson-Walker (FRW) metric
given by
\begin{equation}
ds^2=-dt^2+a^2(t)[dr^2+r^2(d\theta^2+\sin^2{\theta}d\phi^2)],
\end{equation}
where $a(t)$ is the scale factor. We also assume that the spacetime
is filled with only one component fluid i.e. the viscous phantom
energy of energy density $\rho$ (however, in section six, we shall incorporate matter along with phantom energy). The Einstein field equations for
the FRW-metric (in the units $c=1=8\pi G$) are
\begin{equation}
H^2\equiv\Big(\frac{\dot{a}}{a}\Big)^2=\frac{1}{3}\rho,
\end{equation}
and
\begin{equation}
\frac{\ddot{a}}{a}=\dot{H}+H^2=-\frac{1}{6}(\rho+3p),
\end{equation}
where $H$ is the Hubble parameter, $p$ is the effective pressure
containing the isotropic pressure $p_{\text{pe}}$ and the bulk
viscous pressure $p_{\text{vis}}$, given by
\begin{equation}
p=p_\text{pe}+p_{\text{vis}}.
\end{equation}
Here $\rho=\rho_{\text{pe}}+\rho_{\text{vis}}$ and
$p_\text{vis}=-\xi u^\mu_{;\mu}$, where $u^\mu$ is the velocity four
vector and $\xi=\xi(\rho_{\text{vis}},t)$ is the bulk viscosity of
the fluid \cite{ecka}. Eq. (4) shows that negative pressure due to
viscosity contributes in the effective pressure which cause
accelerated expansion. In the FRW model, the expression
$u^\mu_{;\mu}=3\dot{a}/a$ holds. Also, $\xi$ is generally taken to
be positive in order to avoid the violation of second law of
thermodynamics \cite{cata}.

The energy conservation equation is
\begin{equation}
\dot{\rho}+3H(\rho+p)=0.
\end{equation}
Assume that the viscous fluid equation of state (EoS) is
\begin{equation}
p=\omega\rho=(\gamma-1)\rho.
\end{equation}
Note that if $\gamma=0$ (or $\omega=-1$), Eq. (6) represents the EoS
for cosmological constant. Furthermore if $\gamma<0$, it represents
phantom energy. In general, for normal matter $1\leq\gamma<2$.

Using Eqs. (2) - (6), we get the equation governing the evolution of
$H(t)$ for a given $\xi$ as
\begin{equation}
2\dot{H}+3\gamma H^2-3\xi H=0.
\end{equation}
On integration, Eq. (7) gives
\begin{equation}
H(t)=\frac{\exp{\{\frac{3}{2}\int\xi(t)dt\}}}{C+\frac{3}{2}\gamma\exp\{\frac{3}{2}\int\xi(t)dt\}},
\end{equation}
where $C$ is a constant of integration. Note that Eq. (8) can
further be solved to get the evolution of $a(t)$ as
\begin{equation}
a(t)=D\Big(C+\frac{3}{2}\gamma\int\exp\Big\{\frac{3}{2}\int\xi(t)dt\Big\}dt\Big)^\frac{2}{3\gamma},
\end{equation}
where $D$ is a constant of integration. Thus for a given value of
$\xi$ we can obtain expressions of $a(t)$, $\rho(t)$ and $p(t)$
from the system of Eqs. (5) - (9).

\section{Accretion onto black hole}

In the background of FRW spacetime, we consider, as an
approximation, a gravitationally isolated Schwarzschild black hole
(BH) of mass $M$ whose metric is specified by the line element:
\begin{equation}
ds^{2}=-\Big(1-\frac{M}{4\pi r}\Big)dt^{2}+\Big(1-\frac{M}{4\pi
r}\Big)^{-1}dr^{2}+r^{2}(d\theta ^{2}+\sin  ^{2}\theta d\varphi
^{2}). \label{1}
\end{equation}
The background spacetime is assumed to contain one test fluid,
namely the phantom energy with non-vanishing bulk viscous stress
$p_\text{vis}$. The fluid is assumed to fall onto the BH horizon in
the radial direction only which is in conformity with the spherical
symmetry of the BH. Thus, the velocity four vector of the phantom
fluid is $u^{\mu }=(u^{t}(r),u^{r}(r),0,0)$ which satisfies the
normalization condition $u^{\mu }u_{\mu }=-1$. This phantom fluid is
specified by the stress energy tensor for a viscous fluid
\cite{coles,cata}:
\begin{equation}
T^{\mu \nu }=(\rho +p)u^{\mu }u^{\nu }+pg^{\mu \nu }. \label{2}
\end{equation}
Using the energy momentum conservation for $T^{\mu\nu}$, we get
\begin{equation}
ur^{2}M^{-2}(\rho +p)\sqrt{1-\frac{M}{4\pi r}+u^{2}}=C_{1},
\label{3}
\end{equation}
where $u^{r}=u=dr/ds$ is the radial component of the velocity four
vector and $C_{1}$ is a constant of integration. The second constant
of motion is obtained by contracting the velocity four vector of the
phantom fluid with the stress energy tensor $u_{\mu }T_{;\nu }^{\mu
\nu }=0$, which gives
\begin{equation}
ur^{2}M^{-2}\exp \Big[\int\limits_{\rho _{\infty }}^{\rho
_h}\frac{d\rho ^{\prime }}{\rho ^{\prime }+p(\rho ^{\prime
})}\Big]=-A,  \label{4}
\end{equation}
where $A$ is a constant of integration. Also $\rho _{h}$ and
$\rho _{\infty }$ are the energy densities of the phantom fluid at
the horizon of the BH, and at infinity respectively. From Eqs. (12)
and (13) we have
\begin{equation}
(\rho +p)\sqrt{1-\frac{M}{4\pi r}+u^{2}}\exp\Big
[-\int\limits_{\rho _{\infty }}^{\rho _h}\frac{d\rho ^{\prime
}}{\rho ^{\prime }+p(\rho ^{\prime })}\Big]=C_{2}, \label{5}
\end{equation}
with $C_{2}=-C_{1}/A=\rho _{\infty }+p(\rho _{\infty })$. In order
to calculate the rate of change of mass of black hole $\dot{M}$, we
integrate the flux of the bulk viscous phantom fluid over the entire
BH horizon to get
\begin{equation}
\dot{M}=\oint T_t^{r}dS.  \label{6}
\end{equation}
Here $T_t^{r}$ determines the energy momentum flux in the radial
direction only and $dS=\sqrt{-g}d\theta d\varphi $ is the
infinitesimal surface element of the BH horizon. Using Eqs. (12) -
(15), we get
\begin{equation}
\frac{dM}{dt}=\frac{ AM^{2}}{16\pi}(\rho +p), \label{7}
\end{equation}
which clearly demonstrates the vanishing mass of the black hole if
$\rho + p< 0$. Integration of Eq. (16) leads to
\begin{equation}
M=M_0\Big(1-\frac{t}{\tau }\Big)^{-1},  \label{8}
\end{equation}
where $M_0$ is the initial mass of the black hole and modified
characteristic accretion time scale $\tau
^{-1}=[\frac{AM_0}{16\pi}\{(\rho_{\text{pe}} +p_{\text{pe}})-
\frac{3\xi }{t}\ln (\frac{a}{a_0})\}]$, $a_0$ being the value of the
scale factor at time $t_0$. Note that during the integration of (16), we assumed
$\rho_{\text{pe}}$ and $p_{\text{pe}}$ to be constants. In the coming subsections, we shall
take these as time dependent entities.

\section{Accretion of viscous phantom fluid}

We now study the BH mass evolution in two special cases: (a)
constant viscosity; and (b) power law viscosity.

\subsection{Constant bulk viscosity}

For constant viscosity $\xi=\xi_o$, the evolution of $a(t)$ is
determined by using Eq. (9). It gives
\begin{equation}
a(t)=a_0\Big[1+\frac{\gamma H_o
B(t)}{\xi_o}\Big]^\frac{2}{3\gamma},
\end{equation}
where
\begin{equation}
B(t)\equiv \exp{\Big(\frac{3t\xi_o}{2}\Big)}-1.
\end{equation}
Using Eqs. (5), (6) and (8) the density evolution is given by
\begin{equation}
\rho(t)=\frac{\rho_o\exp{(3\xi_o t)}}{\Big[1+\frac{\gamma
H_oB(t)}{\xi_o}\Big]^2}.
\end{equation}
Here $\rho_o=3H_o^2$. Further, for $\gamma<0$ the BR singularity
occurs in a finite time at
\begin{equation}
\tau=\frac{2}{3\xi_o}\ln\Big(1-\frac{\xi_o}{H_o\gamma}\Big).
\end{equation}
Finally, the BH mass evolution is determined by solving Eq. (16) and
(20) to get
\begin{equation}
M=M_0\Big[ 1-\frac{AM_0}{8\pi\gamma}
\Big(\frac{\xi_o}{\Delta}-1\Big)(\xi_o-\gamma H_o) \Big]^{-1},
\end{equation}
where
\begin{equation}
\Delta\equiv\xi_o+(-1+e^{\frac{3t\xi_o}{2}})\gamma H_o.
\end{equation}

This mass is displayed for different values of viscosity at
different times in Table 1.

\begin{tabular}{|l|l|l|l|l|l}\hline\hline
$t\downarrow\xi\rightarrow$&$\xi_1=10^{-17}$&$\xi_2=10^{-18}$&$\xi_3=10^{-19}$&$\xi_4=10^{-20}$
\\ \hline
$t_1=10^{7}$&$3.43427\times10^{-4}$&$2.44662\times10^{-3}$&$6.31285\times10^{-3}$&$7.49184\times10^{-3}$\\
\hline
$t_2=10^{10}$&$3.43544\times10^{-7}$&$2.45261\times10^{-6}$&$6.35248\times10^{-6}$&$7.55357\times10^{-6}$\\
\hline $t_3=10^{13}$&$3.43516\times10^{-10}$&$2.45258\times10^{-9}$&$6.35247\times10^{-9}$&$7.55358\times10^{-9}$\\
\hline $t_4=10^{17}$&$1.23994\times10^{-14}$&$2.10182\times10^{-13}$&$5.86096\times10^{-13}$&$7.01997\times10^{-13}$\\
\hline
\end{tabular}

\vspace{0.09in} \hspace{.25in} Table 1. The mass ratio $M/M_0$ of
black hole for different choices of constant viscosity $\xi_o$. The
initial mass is, throughout, taken to be $50M_\odot$ or $10^{32}$kg.
\vspace{0.09in}

It is apparent from Table.1 that for a fixed viscosity, the mass
ratio decreases with time implying that mass of black hole is
decreasing for an initial mass. Similarly, at any given time, the
mass ratio also decreases with the increase in viscosity. Thus the
greater the value of viscosity parameter, the greater would be its
effects on the BH mass.

\subsection{Power law viscosity}

If the viscosity has power law dependence upon density i.e.
$\xi=\alpha\rho_{\text{vis}}^s$, where $\alpha$ and $s$ are constant
parameters, it has been shown \cite{barr,barr1} that it yields
cosmologies with a BR if $\sqrt{3\alpha}>\gamma$ and $s=1/2$. Thus
we take $\xi=\alpha\rho^\frac{1}{2}$ as a special case. Then the
scale factor evolves as
\begin{equation}
a(t)=a_0\Big(1-\frac{t}{\tau}\Big)^\frac{2}{3(\gamma-\sqrt{3}\alpha)}.
\end{equation}
The density of phantom fluid evolves as
\begin{equation}
\rho(t)=\frac{4}{3\tau^2(\gamma-\sqrt{3}\alpha)^2}\Big(1-\frac{t}{\tau}\Big)^{-2},
\end{equation}
or in terms of critical density $\rho_{\text{cr}}$ as
\begin{equation}
\rho(t)=\rho_{\text{cr}}\Big(1-\frac{t}{\tau}\Big)^{-2}.
\end{equation}
The corresponding BR time $\tau$ is given by
\begin{equation}
\tau=\frac{2}{3(\sqrt{3}\alpha-\gamma)}H_o^{-1}.
\end{equation}
Finally, the mass evolution of BH is determined by using Eq. (16)
and (25) is
\begin{equation}
M=M_0\Big[1+\frac{AM_0}{4\pi(\sqrt{3}\alpha-\gamma)}\frac{t}{\tau(\tau-t)}
\Big]^{-1}.
\end{equation}
Note that when $\alpha=0$, this case reduces to that of Babichev
et al \cite{babi}. The mass in (28) in displayed for different
values of EoS parameter $\gamma$ at different times in Table 2 and
displayed graphically in Figure 1. As shown, the mass decreases
gradually with the decrease in the EoS parameter $\gamma$. Note that
we have not graphically displayed the mass for different viscosities
given in Table 1 because the variation is not significantly
different for most time scales.

\begin{tabular}{|l|l|l|l|l|l}\hline\hline
$t\downarrow\gamma\rightarrow$&$\gamma_1=-1\times10^{-1}$&$\gamma_2=-2\times10^{-1}$&$\gamma_3=-3\times10^{-1}$&$\gamma_4=-4\times10^{-1}$
\\ \hline
$t_1=10^{10}$&$4.71915\times10^{-5}$&$2.35963\times10^{-5}$&$1.5731\times10^{-5}$&$1.17983\times10^{-5}$\\
\hline
$t_2=10^{13}$&$4.79136\times10^{-8}$&$2.35968\times10^{-8}$&$1.57312\times10^{-8}$&$1.17984\times10^{-8}$\\
\hline $t_3=10^{17}$&$4.66492\times10^{-12}$&$2.30523\times10^{-12}$&$1.51867\times10^{-12}$&$1.12539\times10^{-12}$\\
\hline $t_4=10^{20}$&$4.97349\times10^{-14}$&$5.20946\times10^{-14}$&$5.28811\times10^{-14}$&$5.32744\times10^{-14}$\\
\hline
\end{tabular}

\vspace{0.09in} \hspace{.25in} Table 2. The mass ratio $M/M_0$ of
black hole for different choices of equation of state. The initial
mass is $50M_\odot$ or $10^{32}$kg.
\vspace{0.09in}

\section{Examples}
We now solve examples to demonstrate the accretion of viscous
phantom energy onto a BH. The formalism is adapted from \cite{babi}.

\subsection{Viscous linear EoS}

We choose the viscous linear EoS, $p=\omega\rho_{\text{pe}}-3H\xi_o$
with $\omega<-1$. The ratio of the number densities of phantom fluid
particles at the horizon and at infinity is given by
\begin{equation}
\frac{n(\rho^{\text{pe}} _h)}{n(\rho^{\text{pe}}_\infty
)}=\Big[\frac{\rho^{\text{pe}} _h(1+\omega )-3\xi_o H}{\rho
^{\text{pe}}_{\infty }(1+\omega )-3\xi_o H}
\Big]^{\frac{1}{(1+\omega )}}.
\end{equation}
The critical points of accretion (the point where the speed of fluid
flow becomes equal to the speed of sound i.e. $u_\ast^2=c_s^2$) are
given by
\begin{equation}
u_{\ast }^{2}=\frac{\omega }{1+3\omega };\ \ x_{\ast
}=\frac{1+3\omega }{2\omega }.
\end{equation}
The constant $A$ appearing in Eq. (16) is determined to be
\begin{equation}
A=\frac{|1+3\omega|}{4|\omega |^{3/2}}^{\frac{1+\omega }{2\omega }}.
\end{equation}
Notice that the constant $A$ is the same as for the non-viscous case
\cite{babi}. Also, the density of phantom energy at the horizon is
given by
\begin{equation}
\rho^{\text{pe}} _{h}=\frac{3\xi_o H}{1+\omega
}+\Big(\frac{4}{A}\Big)^{ \frac{\omega -1}{\omega +1}}\Big(\rho
_{\infty }-\frac{3\xi_o H}{1+\omega }\Big).
\end{equation}
Moreover, the speed of flow at the horizon is
\begin{equation}
u_{h}=-\Big(\frac{A}{4}\Big)^{\frac{\omega }{(\omega +1)}}.
\end{equation}
The speed is negative as it is directed towards the BH. Also, the
characteristic evolution time scale of the BH is given by
\begin{equation}
\tau ^{-1}=4\pi M_0\frac{(1+3\omega )}{4\omega
^{3/2}}^{\frac{1+\omega }{2\omega }} \Big\{\rho^{\text{pe}}_\infty(
1+\omega)- \frac{3\xi_o }{t}\ln \Big(\frac{a}{a_0}\Big)\Big\}.
\end{equation}
Finally, substituting Eq. (34) in (17) we get the mass evolution of
a BH in bulk viscous cosmology
\begin{equation}
M=M_0\Big[1-4\pi M_0t\frac{(1+3\omega )}{4\omega
^{3/2}}^{\frac{1+\omega }{2\omega }} \Big\{\rho^{\text{pe}}_\infty(
1+\omega)- \frac{3\xi_o }{t}\ln\Big
(\frac{a}{a_0}\Big)\Big\}\Big]^{-1}.
\end{equation}
Since $\rho^{\text{pe}}_{\infty}$ is unknown for our purpose, we
have not evaluated $M$ for different times numerically for tabular
and graphical presentation.

\subsection{Viscous non-linear EoS}

We here choose the EoS, $p=\omega\rho_{\text{pe}}-3H\xi(\rho_{\text{vis}})$ with
$\omega<-1$, where $\xi(\rho_{\text{pe}})=\alpha\rho_{\text{pe}}^s$ with $\alpha$ and $s$
are constants. The ratio of number densities is given by
\begin{equation}
\frac{n(\rho^{\text{pe}} _h)}{n(\rho^{\text{pe}}_\infty
)}=\Big(\frac{\rho_h}{\rho_\infty}\Big)^{\frac{s}{(s-1)(1+\omega)}}\Big(\frac{\rho_\infty(1+\omega)-3H\alpha\rho_\infty^s}{\rho_h(1+\omega)-3H\alpha\rho_h^s}\Big)
\end{equation}
The constant $A$ appearing in Eq. (16) is determined to be
\begin{equation}
A=\Big|\Big(\frac{\rho_h}{\rho_\infty}\Big)^{\frac{2s}{(s-1)(1+\omega)}}\Big(\frac{\rho_\infty(1+\omega)-3H\alpha\rho_\infty^s}{\rho_h(1+\omega)-3H\alpha\rho_h^s}\Big)^3\Big|.
\end{equation}
The speed of flow at the horizon becomes
\begin{equation}
u_h=-\Big(\frac{\rho_h}{\rho_\infty}\Big)^{\frac{2s}{(s-1)(1+\omega)}}\Big(\frac{\rho_\infty(1+\omega)-3H\alpha\rho_\infty^s}{\rho_h(1+\omega)-3H\alpha\rho_h^s}\Big)^2.
\end{equation}
The critical points of accretion are given by
\begin{equation}
u_\ast^2=\frac{\omega-3s\alpha\rho_h^{s-1}}{1+3(\omega-3s\alpha\rho_h^{s-1})};\
\
x_\ast=\frac{1+3(\omega-3s\alpha\rho_h^{s-1})}{2(\omega-3s\alpha\rho_h^{s-1})}.
\end{equation}
The characteristic evolution time scale $\tau$ is given by
\begin{equation}
\tau=\Big[4\pi
M_0\Big(\frac{\rho_h}{\rho_\infty}\Big)^{\frac{2s}{(s-1)(1+\omega)}}\Big(\frac{\rho_\infty(1+\omega)-3H\alpha\rho_\infty^s}{\rho_h(1+\omega)-3H\alpha\rho_h^s}\Big)^3\Big\{
\rho_\infty(1+\omega)-3\frac{\alpha\rho^s_\infty}{t}\ln{\Big(\frac{a}{a_0}\Big)}\Big\}
\Big]^{-1}.
\end{equation}
Finally, using Eqs. (40) in (17),  the BH mass evolution is given by
\begin{eqnarray}
M&=&M_0\Big[1-4\pi
M_0t\Big(\frac{\rho_h}{\rho_\infty}\Big)^{\frac{2s}{(s-1)(1+\omega)}}\Big(\frac{\rho_\infty(1+\omega)
-3H\alpha\rho_\infty^s}{\rho_h(1+\omega)-3H\alpha\rho_h^s}\Big)^3\nonumber
\\&\;& \times\Big\{
\rho_\infty(1+\omega)-3\frac{\alpha\rho^s_\infty}{t}\ln{\Big(\frac{a}{a_0}\Big)}\Big\}\Big]^{-1}.
\end{eqnarray}
As before, $\rho_\infty$ is unknown, but further $\rho_h$ is also
unknown. As such, we again do not provide a tabular or graphical
presentation.

\section{Black holes accreting both matter and viscous phantom fluid}

We now consider a two component fluid, the viscous dark energy and
matter. The matter part may be composed of both baryonic and
non-baryonic matter. It is taken to be a perfect fluid while the PE
is taken as a bulk viscous fluid. The effective pressure is
represented by Eq. (4). The corresponding Einstein field equations
(EFE) for the two component fluid become:
\begin{equation}
R_{\mu \nu }-\frac{1}{2}g_{\mu \nu }=T_{\mu \nu }+T_{\mu \nu
}^{\text{m}}.
\end{equation}%
The stress-energy tensor representing the two component fluid is
given by
\begin{equation}
T^{\mu \nu }=(\rho+p+\rho _\text{m})u^{\mu }u^{\nu}
+pg^{\mu \nu }.
\end{equation}
Here $\rho _{\text{m}}$\ is the energy
density of the pressureless matter. Energy conservation holds
independently for both fluids:
\begin{equation}
\dot{\rho }+3H(\rho+p)=0,
\end{equation}
\begin{equation}
\dot{\rho }_{\text{m}}+3H\rho _{\text{m}}=0.
\end{equation}
Integrating Eq. (45), we have
\begin{equation}
\rho _{\text{m}}=\rho _{\text{m}_0}a^{-3},
\end{equation}
where $\rho_{\text{m}_0}=\rho_\text{m}(t_0)$. Similarly, integrating Eq. (44)
leads to
\begin{equation}
\rho=\rho_{\text{m}}\Big[\Big(\Xi+\frac{K}{3}a^{3/2}\Big)^{2}
-1\Big],
\end{equation}
where $\Xi$ is a constant and $K$ is given by
\begin{equation}
K=\frac{3\sqrt{3}\xi_o}{\sqrt{\rho_{\text{m}_0}}},
\end{equation}%
Thus the total energy density of the two component fluid is given by
\cite{coli}
\begin{equation}
\rho \equiv \rho+\rho _{\text{m}}=\rho _{\text{m}_0 }a^{-3}\Big(
\Xi+\frac{K}{3}a^{3/2} \Big)^{2}.
\end{equation}%
Using Eqs. (45) in (16) the evolution of black hole mass is given by
\begin{eqnarray}
M&=&M_0\Big[ 1-4\pi A M_0\Big[\frac{\gamma\rho_{\text{m}_0}}{H(t)}\Big\{
\frac{K^2}{9}\ln{\Big(\frac{a}{a_0}\Big)}-\frac{\Xi}{9a^3}
(3\Xi+4a^{3/2}K)\nonumber
\\&\;&+\frac{\Xi}{9a_0^3}(3\Xi+4a_0^{3/2}K)
\Big\}\Big] \Big]^{-1},
\end{eqnarray}
where the scale factor $a(t)$ evolves as
\begin{equation}
a(t)=\Big[ \frac{3}{K}(e^{\frac{K}{2}\sqrt{\rho_{\text{m}_0}/3}t+D_1}-\Xi)
\Big]^{2/3},
\end{equation}
and $D_1$ is the constant of integration determined by choosing
$t=0$ to get
\begin{equation}
D_1=\frac{2}{K}\ln{\Big(\frac{K}{3}a_0^{3/2}+\Xi\Big)}.
\end{equation}
As pointed out in the next section, we cannot correctly discuss a BR
scenario. However we can take a spacetime approximating it
sufficiently earlier than the BR. We can than see its asymptotic
behavior. when the scale factor shoots to infinity, the three terms
in Eq. (50) will contribute significantly in the BH mass evolution.
The mass will decrease by the accretion of PE ($\gamma<0$) due to
its strong negative pressure and is manifested in Eq. (50). Notice
that the final expression for BH mass depends only on the initial
matter density $\rho_{m_0}$ in addition to constant bulk viscosity
$\xi_o$. The corresponding behavior of BH mass evolution is shown in
Figures 2 and 3 for different values of model parameters. Thus for a
shift of parameter $\gamma$ by 2, yields in the decline of mass
ratio by a factor of 2. The decline in the mass of the BH is
observed with time showing that phantom energy accretion will be
dominant over matter accretion.

\section{Conclusion and discussion}

We have analyzed the accretion of bulk viscous phantom energy onto a
BH. The modeling is based on the relativistic model of accretion for
compact objects. The viscosity effects in cosmology are used to give
an alternative to cosmic accelerated expansion other then dark
energy and quintessence. The evolution of BHs in such a Universe
accreting viscous phantom energy would result in a gradual decrease
in mass. This gradual decline would be faster than the non-viscous
case \cite{babi} due to additional terms containing viscosities
coupled with mass. Lastly, it is shown that BHs accreting both matter
and viscous PE will also meet with the same fate as the viscous
forces dominate over the matter component for sufficiently large
scale factor $a(t)$.

From this analysis, we can draw the conclusion that PE containing
viscous stresses can play a significant role in the BH mass
evolution if the viscosity is sufficiently high for an appropriate
EoS. Though the viscous stresses are negligibly small
$O(10^{-8}Nsm^{-2})$ at the local scale of space and time they can
play a significant role in time scales of $\sim$ Gyrs. The higher
the viscosity of the phantom fluid, the sharper the decrease in the
BH mass. BHs of all masses, ranging from the solar mass to the
intermediate mass to the supermassive, will all meet the same fate.

As an extension to this problem, it is interesting to study the
accretion of the phantom fluid onto primordial BHs that had formed
due to initial density fluctuations in the primordial plasma. The
mini-primordial BHs evaporating now via Hawking radiation would
have a different initial mass and hence abundance than the standard
scenario expects. This work is reported in a separate paper
\cite{mq}.

Notice that we have used the Friedmann model which is represented by
an asymptotically curved spacetime and at the same time the
Schwarzschild black hole, which is asymptotically flat. This may
seem contradictory. Schwarzschild black hole has been dealt with in
the context of closed Friedmann cosmology \cite{aq1,aq2,aq3}. Any
global problem in approximating the full situation by a
Schwarzschild black hole inserted into Friedmann model arise near
the big bang or the big crunch, defined in terms of the york time
\cite{mtw} as shown elsewhere \cite{aq5}, the effect will be at
extremely late times in terms of the usual time parameter. More
complete analysis of the asymptotic behavior near a singularity is
also available \cite{aq6}, as such if we stay near to a singularity
in spacetime, the approximation will be extremely good. Consequently
our analysis will be satisfactory for black holes formed well after
the big bang greater then $10^{-40}$s and of the Big Rip (presumably
much more before $10^{-40}$s the rip). It is clear that we are
unable to say whether there would/would not be a Big Rip as our
analysis excludes it.

\subsubsection*{Acknowledgment} MJ would like to thank M. Akbar for
useful discussions during the work. AQ is grateful to AS-ICTP,
Trieste, Italy for travel support and to INFN and the Department of
Physics of Salento University at Lecce, for local hospitality.

\newpage
 \pagebreak
 \begin{figure}
\includegraphics{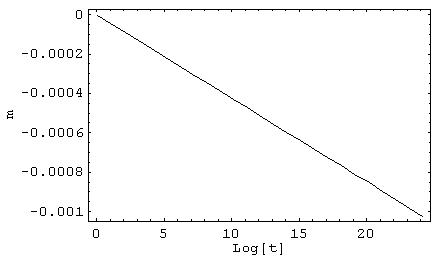}\\
\caption{ For an initial mass of black hole $M_0=10^{32}$kg, the
evolution of the mass parameter $m=M/M_0-1$ is plotted against the
logarithmic time with $\alpha=10^{-5}$ and $t_H=10^{17}s$.}
\end{figure}
\begin{figure}
\includegraphics{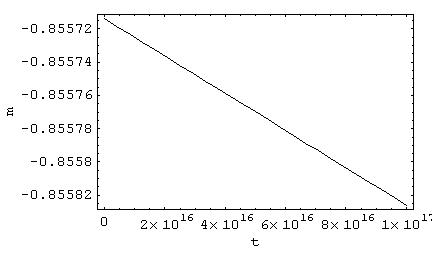}\\
\caption{ For an initial mass of black hole $M_0=10^{32}$kg, the
evolution of $m$ is plotted against the time parameter $t$ with
$A=1/3$, $\Xi=3$, $\xi_o=10^{-16}kgm^{-1}s^{-1}$ and
$\gamma=-10^{-1}$ while $H\approx
2.33\times10^{-18}$m.}
\end{figure}
\begin{figure}
\includegraphics{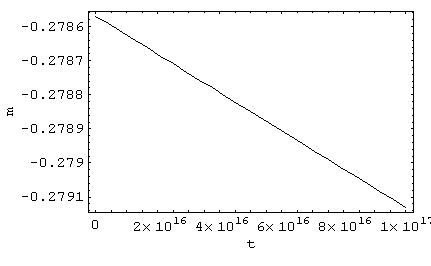}\\
\caption{ For an initial mass of black hole $M_0=10^{32}$kg, the
evolution of $m$ is plotted against the time parameter $t$ with
$A=1/3$, $\Xi=3$, $\xi_o=10^{-16}kgm^{-1}s^{-1}$ and
$\gamma=-2\times10^{-1}$ while
$H\approx 2.33\times10^{-18}$m.}
\end{figure}

\pagebreak

\newpage
\end{document}